\documentclass[aps,prb,twocolumn,floatfix,amsmath,amssymb]{revtex4}
\usepackage{graphicx}
\usepackage{dcolumn}% Align table columns on decimal point
\usepackage{bm}% bold math
\usepackage{epsfig,psfrag,amsmath,amssymb,float}
\input{epsf}
\newcommand{\vv}[1]{{\bf #1}}

\begin{document}

\title{Electron-phonon vertex in the two-dimensional one-band Hubbard model} 
\author{Z.~B.~Huang, W.~Hanke, and E.~Arrigoni}
\affiliation{Institut f\"ur Theoretische Physik, Universit\"at W\"urzburg,
am Hubland, 97074 W\"urzburg, Germany}
\author {D.~J.~Scalapino}
\affiliation{Department of Physics,
University of California,
Santa Barbara, California 93106-9530 USA}

\date{\today}
%\maketitle
%\vskip 0.0truecm

\begin{abstract}
Using quantum Monte Carlo techniques,  we study the effects of 
electronic correlations on the effective electron-phonon (el-ph) 
coupling in a two-dimensional one-band Hubbard model. 
We consider a momentum-independent bare ionic el-ph coupling.
In the weak- and intermediate-correlation regimes, we find that the 
on-site Coulomb interaction $U$ acts to effectively suppress the ionic
el-ph coupling at all electron- and phonon- momenta.
In this regime, our numerical simulations are in good agreement with 
the results of perturbation theory to order $U^2$. However, 
entering the strong-correlation regime, we find that the forward 
scattering process stops decreasing and 
begins to substantially increase as a function of $U$, leading to an 
effective el-ph coupling which is peaked in the forward direction. 
Whereas at weak and intermediate Coulomb interactions, screening
is the dominant correlation effect suppressing the el-ph coupling,
at larger $U$ values irreducible vertex corrections become more
important and give rise to this increase. These vertex corrections
depend crucially on the renormalized electronic structure of 
the strongly correlated system.
\end{abstract}


\maketitle

%\narrowtext
The role of the el-ph interaction in the physics of the high
$T_c$ cuprate superconductors remains unclear.  On the one hand, the linear
$T$ dependence of the resistivity up to high temperatures and the small
value of the isotope coefficient of the optimally doped materials suggest
that the el-ph interaction plays a secondary role \cite{IFT98}.
The fact that the undoped cuprates are Mott antiferromagnetic
insulators supports the notion that strong Coulomb interactions are
dominant and that the essential physics is contained in the Hubbard and t-J
models \cite{And02}. On the other hand, however, a variety of experiments
also display
pronounced phonon and electron-lattice effects in these materials: 
superconductivity-induced phonon renormalization~\cite{raman}, 
large isotope coefficients away from optimal doping~\cite{iso1}, 
tunneling phonon structures~\cite{tunnel1}, etc., give evidence of 
significant el-ph coupling.  Recently, photoemission data indicated 
a sudden change in the electron dispersion near a characteristic
energy scale~\cite{lanzara}, which is possibly caused by coupling of 
electronic quasiparticles to phonon modes. 

To elucidate the effects of strong electronic correlations on the el-ph
interaction, several authors have calculated the el-ph vertex function 
in the one- and three-band Hubbard models based on $(1/N)$ expansion within 
slave-boson~\cite{kim,grilli} and X operator~\cite{ZK96} formalisms. 
One finding \cite{ZK96} is that for the ionic, i.e.~onsite, 
el-ph coupling
in the underdoped regime, 
the backward scattering with large phonon momentum transfer
 is suppressed much more than 
the forward scattering with small phonon momentum transfer. 
Based on this finding,
it was argued that a forward peaking of the renormalized el-ph
vertex could account for the absence of phonon features in the transport
data.  In addition, an el-ph interaction which is peaked at 
small phonon momentum transfer contributes to an
attractive interaction in the $d_{x^2-y^2}$-pairing channel
\cite{ZK96,BS96}.
One should notice that these previous calculations were limited by the
approximate nature of $1/N$ and slave-boson treatments, moreover, they
were carried out for $U\rightarrow\infty$.

Here, we would like to use a more accurate numerical method
to gain further insight into the way in which strong
electron correlations dress the el-ph coupling for a range
of values of $U$ from weak to strong correlation.
This analysis is important since it turns out that the $U$ dependence
of the el-ph coupling shows a dramatic change as a function
of $U$ (see below).
Specifically, we apply the
determinantal Monte Carlo \cite{BSS81} algorithm to
find the single-particle response to external phonon fields
in the one-band Hubbard model.
In particular, we will calculate an effective el-ph coupling~\cite{pq} 
$g(p, q)$ (effective el-ph vertex) 
for scattering quasiparticles near the Fermi surface (which includes
screening, vertex corrections, and the quasiparticle renormalization)
produced by the Hubbard $U$.

\begin{center}
\begin{figure}
\epsfig{file=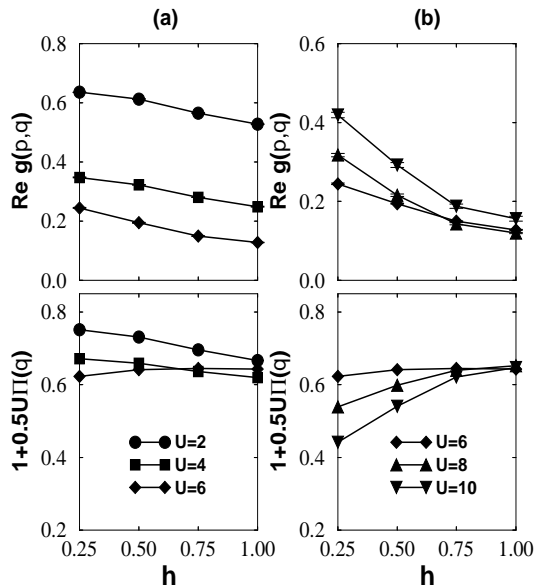,height=7cm,width=8cm,angle=270}
\caption{Real part of the effective el-ph coupling $g(p, q)$
and the polarization factor $1+{1\over 2}U\Pi$ versus
${\vv q}$ for (a) $U \le 6$ and (b) $U \ge 6$. Here
${\vv q}=\pi(h,h)$ with $h$ the tick label of the $x$ axis. 
The incoming electron carries momentum ${\vv p}=(-\pi,\,0)$ and
the value of $U$ is indicated by the shape of the symbol.}
\label{Gamma}
\end{figure}
\end{center}

Our principle findings are: 
(1) Initially (up to the value of $U\approx 6t$), 
as the Hubbard-Coulomb interaction $U$ increases, 
the el-ph coupling is {\it suppressed} by electronic correlations
for all phonon and electron momenta, and in particular for the backward 
scattering around $\vv q=(\pi,\,\pi)$. The suppression is due to
the conventional screening term.
(2) The behavior changes even qualitatively in the strong-correlation 
regime ($U\ge 6t$). Here, the effective el-ph coupling at {\it small}
phonon momentum transfer {\it increases} with {\it increasing} $U$, 
while the one at large phonon momentum transfer 
appears to saturate (see Fig.~\ref{Gamma}). 
The increase of the el-ph coupling in the forward direction is due to
irreducible vertex corrections, which become the dominant correlation
effects at strong Coulomb interactions. These vertex corrections are
intimately connected with the renormalized electronic structure.
It is argued that the picture of a ``spin-bag" quasiparticle can explain the 
qualitatively different el-ph couplings for small and large phonon momenta.
Furthermore, we would like to stress that our numerical results 
for the charge compressibility, which decreases 
monotonically with increasing $U$, rule out the explanation of 
the increase of the el-ph vertex at small ${\vv q}$ as a function
of  $U$ in terms of a close-by phase separation or charge 
instability~\cite{cerruti}.

Our starting point is the one-band Hubbard model,
\begin{eqnarray}
\label{ham}
 H = -t \sum_{\langle ij \rangle,\sigma}
     (c_{i\sigma}^\dagger c_{j\sigma}^{\,}
     +c_{j\sigma}^\dagger c_{i\sigma}^{\,})
     + U \sum_i n_{i\uparrow}n_{i\downarrow},
\end{eqnarray}
The operators $c_{i\sigma}^\dagger$ and $c_{i\sigma}^{\,}$  as usual create and 
destroy an electron with spin $\sigma$ at site $i$, respectively and the sum
$\langle ij\rangle$ is over nearest-neighbor lattice sites.  Here, $U$ is the onsite 
Coulomb interaction and we will choose the nearest-neighbor hopping $t$ as
the unit of energy.

\begin{center}
\begin{figure}
\epsfig{file=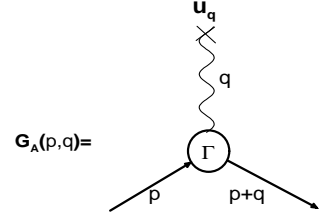,height=4.0cm,width=3.0cm,angle=270}
\medskip
\caption{Diagrammatic representation of $G_A(p,q)$ 
within linear response to $u_{\vv q}$. The thick solid lines 
represent dressed single-particle Green's functions of 
the Hubbard model. The wavy line denotes the external 
perturbation in Eq.~\eqref{el-ph}.}
\label{vertex}
\end{figure}
\end{center}

In our simulations, we have used the linear-response technique
% to an
%imaginary-time dependent perturbation~\cite{fint} 
in order to extract the 
el-ph vertex function. 
In this method, one formally adds to Eq.~\eqref{ham} the interaction
with a momentum- and (imaginary) time-dependent lattice distortion (phonon) field
$u_{ \vv q} e^{-i q_0 \tau}$ in the form~\cite{pq,fint}
\begin{equation}
\label{el-ph}
H_{el-ph}= \sum_{\vv k \vv q\sigma} g_{\vv k\vv q}^{0}
\ c_{\vv k+\vv q\sigma}^{\dagger}c_{\vv k\sigma} \
u_{ \vv q} \ e^{-i q_0 \tau}\;,
\end{equation}
where $g_{kq}^{0}$ is the bare el-ph coupling. 
One then considers the 
%%response of the
``anomalous'' single-particle propagator in the presence of this
perturbation defined as~\cite{pq}
\begin{equation}
\label{gq}
G_{A}(p,q)\equiv -\int_{0}^{\beta}d\tau\ e^{i(p_{0}+q_{0}) \tau}
 \langle T_{\tau}c_{\vv p+\vv q\sigma}(\tau)c_{\vv p\sigma}^{\dagger}
(0)\rangle_{H+H_{el-ph}},
\end{equation}
where $\langle\rangle_{H+H_{el-ph}}$ is Green's function
evaluated with the Hamiltonian $H+H_{el-ph}$. 
 Diagrammatically $G_A(p, q)$ has
the structure shown in Fig.~\ref{vertex} 
so that 
the el-ph 
vertex function
$\Gamma(p,q)$ 
 can be expressed quite generally in
terms of $G_A$ and of the single-particle Green's function $G(p)$ in
the form
\begin{equation}
\Gamma (p, q) = \lim_{u_{\vv q}\to 0} \frac{1}{u_{\vv q} }\frac{  G_A (p, q)}{G(p+q)\, G (p)
  }  \;,
\label{five}
\end{equation}
It is, thus, sufficient to calculate
the leading linear response of $G_A$ to $H_{\rm el-ph}$, which is given by
\begin{eqnarray}
G_A(p,q) = u_{\vv q} \ 
 \int_{0}^{\beta} d\tau e^{i(p_{0}+q_{0})\tau} 
\int_{0}^{\beta} d\tau^{'} e^{-i q_{0} \tau'}
\sum_{\vv k\vv q\sigma^\prime}g_{\vv k\vv q}^{0} 
\times \nonumber\\
 \langle T_{\tau}c_{\vv k+\vv q\sigma^\prime}^{\dagger}(\tau'+0^{+})c_{\vv
  k\sigma^\prime}(\tau')
 c_{\vv p+\vv q\sigma}(\tau)c_{\vv p\sigma}^{\dagger}(0)\rangle_{H},
\label{four}
\end{eqnarray}
where $0^+$ is a positive infinitesimal.
The two-particle Green's function in Eq.~\eqref{four} is evaluated 
with respect to the pure Hubbard Hamiltonian  (Eq.~\eqref{ham}).
Since we have only considered the linear-response contribution 
from  the phonon field, the el-ph vertex $\Gamma$ contains full
contributions from Coulomb interactions only~\cite{ref1},

Close to the Fermi energy, the single-particle Green's function 
can be written as
\begin{equation}
G(p) = \frac{1}{Z(p) (i\ p_0 - E_{\vv p})},
\label{seven}
\end{equation}
where $Z(p)$ is the wave-function renormalization and
$E_{\vv p}$ the quasiparticle excitation energy~\cite{ref2}.
Then for electron scattering processes which involve states near the Fermi
surface, the effective el-ph coupling reads
\begin{equation}
g(p, q) = \frac{\Gamma(p, q)}{\sqrt{Z(p)\, Z(p+q)}},
\label{eight}
\end{equation}

In the following, we will focus on the case of an {\it ionic}
el-ph coupling,
in which the bare coupling $g_{\vv p\vv q}^{0}$ is a constant $g^{0}$. 
Since we are considering linear terms in $g^{0}$ only, we can
set $g^{0}$ equal to 1.
This corresponds to the simple Holstein form of the 
el-ph interaction, which is an important limiting case. 
Moreover, having the bare interaction
independent of $\vv p$ and $\vv q$ makes it easier to see 
modifications, which arise from the strong $U$ correlation effects. 

\begin{center}
\begin{figure}
\epsfig{file=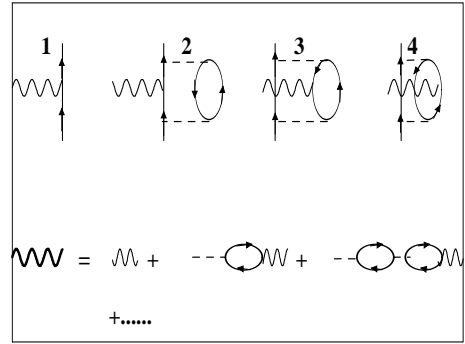,height=6.0cm,width=4.5cm,angle=270}
\medskip
\caption{Low-order Feynman diagrams for the irreducible el-ph
vertex $\Lambda(p,q)$ (top) and low-order polarization graphs (lower) that enter
the full vertex $\Gamma$. The thin solid lines are the non-interacting
Green's functions and the dashed lines represent the Hubbard interaction $U$.
The thin (thick) wavy line stand for the bare (screened phonon) fields.
}
\label{Diagram}
\end{figure}
\end{center}

The low order $U$ and $U^2$ vertex contributions to $\Gamma$ are displayed in 
Fig.~\ref{Diagram}. The diagrams shown at the bottom of Fig.~\ref{Diagram}
are the leading terms of the RPA approximation 
$(1- \frac{1}{2}\, U\,\Pi_0 (q))^{-1}$
to the polarization correction, with $\Pi_0(q)$ the contribution from the
single bubble.  The exact Monte Carlo result for the polarization
correction is $1+ \frac{1}{2}\, U\, \Pi (q)$ with
\begin{eqnarray}
\Pi (q) & = & - \int^\beta_0 d\tau\ e^{-i\ q_0 \tau}
\ \left\langle T_\tau \rho_{\vv q} (\tau) \rho^\dagger_{\vv q} (0)\right\rangle,
\nonumber\\
\noalign{\hbox{and}}
\rho^\dagger_{\vv q} & = & {1\over\sqrt{N}}\sum_{\vv k\sigma}
c_{\vv k+\vv q\sigma}^{\dagger}c_{\vv k\sigma},
\label{ten}
\end{eqnarray}
With this in mind, $\Gamma$ can be written in terms of the screening factor
and an irreducible vertex $\Lambda(p,q)$, which is the sum of graphs
that can not be separated into two pieces by cutting a single dashed 
Coulomb interaction line $U$ (see Fig.~\ref{Diagram}), \textit{i.e.}
\begin{equation}
\Gamma(p,q)=(1+{1 \over 2}U\Pi(q))\Lambda(p,q),
\label{eleven}
\end{equation}

Thus, the strong-correlation effects associated with the Hubbard-Coulomb
interaction $U$ lead to an effective el-ph coupling which can be
expressed in the canonical form
\begin{equation}
g(p, q) = \frac{\left(1+\frac{1}{2} U \Pi (q)\right) \Lambda(p, q)}{(Z(p)
Z(p+q))^{\frac{1}{2}}}.
\label{twelve}
\end{equation}
and one sees that it consists of a product of an irreducible vertex
$\Lambda
(p, q)$, a screening factor $(1+{1\over 2} U\Pi(q))$, and a quasiparticle
renormalization $(Z(p) Z(p+q))^{-{1\over 2}}$ factor.

Our numerical Monte Carlo simulations were performed on an 
$8 \times 8$ lattice at an inverse temperature
$\beta=2$ and a filling $<n>=0.88$. 
We have set the frequencies to their minimum values, i.e.,
$p_0=\pi T$ for fermions and $q_0=0$ for bosons.
We have checked some special cases for which one can reach lower
temperatures, namely a 2D system at weak correlation and/or
with large doping ($<n>=0.65$).
For these systems, we have found that the real part of the vertex function 
$\Gamma((\pi T,{\vv p}),(0,{\vv q}))$ depends only weakly on temperature, 
and the imaginary part always vanishes as $T \rightarrow 0$.
In the following we will, therefore, focus on the real part of the 
vertex function at $p_0=\pi T$.
Comparison with exact diagonalization on a 4-site ring demonstrates 
that the difference in $\Gamma(p,q)$ between the two results
is less than two percent up to $U=8$.

We are interested in el-ph scattering processes in which
the incoming and the outgoing electron momenta ${\vv p}$ and ${\vv
  p+\vv q}$
are close to the Fermi surface. For an $8 \times 8$ lattice doped near
half-filling, the $q$ and $U$ dependence of $g(p,q)$ for the
scattering processes on the half-filled diamond Fermi surface is 
studied. In particular, we will examine initial states corresponding to 
${\vv p}=(-\pi,\,0)$ and ${\vv p}=(-\pi/2,\,\pi/2)$. 
Other choices of ${\vv p}$ and ${\vv p+\vv q}$ close to 
the half-filled diamond Fermi surface give qualitatively similar results to
those reported here.

Monte Carlo results for $g(p, q)$ and for the polarization factor 
$(1+{1\over 2}U\Pi)$ are shown in Fig.~\ref{Gamma}. The left hand side,
Fig.~\ref{Gamma}a, shows the behavior in the weak- and 
intermediate-correlation regimes.
The right hand side of the figure, Fig.~\ref{Gamma}b, 
shows similar results when the system enters the strong-correlation regime.
One can clearly see that when the Hubbard $U$ is smaller than $U\approx 6$,
$g(p, q)$ {\it decreases} as a function of $U$ from its bare value 
$g^{0}=1$, for all momentum transfers. 
Then, as the interaction $U$ increases to the strong-correlation
case ($U \sim W=8t$), the effective el-ph coupling
begins to {\it increase}. This behavior is particularly evident 
at smaller values of momentum transfer. 
Our finding at large phonon momentum is similar to that of Deppeler
{\it et al.}'s work~\cite{holdmft}, which shows that the local el-ph 
interaction is suppressed by both electronic correlations and dynamic phonon
vertex corrections.
In the strong-correlation regime, the overall $\vv q$ dependence of the 
el-ph coupling agrees reasonably well with the results of
the $1/N$ expansion~\cite{ZK96} which are obtained for the 
$U\rightarrow\infty$ limit. 
However, in our case the interesting behavior 
is that the effective el-ph coupling as a function of $U$ is
nonmonotonic, first decreasing and then, at physically interesting
values of $U$, increasing. 
This finding deviates from the prediction 
of a Fermi-liquid  analysis~\cite{grilli}. According to this analysis,
$\lim_{\vv q \to 0,q_0 = 0} \ g(p,q)\propto\frac{1}{1+F_{0}^{s}}$
with $F_{0}^{s}$ the zero-harmonic symmetric Landau amplitude so that
the effective el-ph coupling decreases monotonically
with increasing $U$ since $F_{0}^{s}$ becomes larger with $U$
(except when approaching a charge instability).

From Fig.~\ref{Gamma}, one can see that the polarization factor
acts quite generally to suppress the el-ph coupling.
At large momentum transfer, this quantity saturates as $U$ increases, 
while at small momentum transfer it continues to decrease.
Comparison between $g(p,q)$ and the polarization factor
indicates that at weak correlation screening is the dominant 
correlation effect suppressing the el-ph coupling.
On the other hand, with an increasing Hubbard $U$, 
the vertex corrections $\Lambda$ become more
important, making the effective el-ph coupling peaked
in the forward scattering direction. 

\begin{center}
\begin{figure}
\epsfig{file=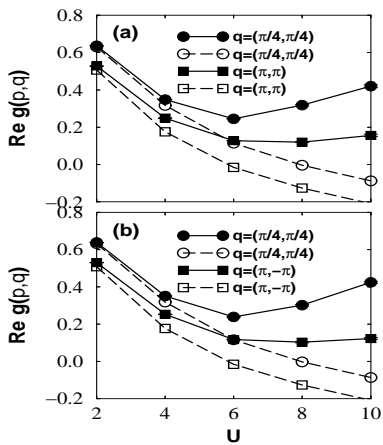,height=5cm,width=6cm,angle=270}
\caption{Real part of $g(p, q)$ as a function of $U$ 
for (a) ${\vv p}=(-\pi,\,0)$ and (b) ${\vv p}=(-\pi/2,\,\pi/2)$.
The value of $\vv q$ is indicated by the shape of the symbol.
The solid circles are Monte Carlo results and the open symbols show 
the perturbation theory contributions shown in Fig.~\ref{Diagram}. 
}
\label{MC_Pert}
\end{figure}
\end{center}

In order to see the $U$ dependence more clearly, in Fig.~\ref{MC_Pert}
Quantum Monte Carlo calculations are compared with perturbation theory
for different values of $U$. 
Here, the solid symbols are Monte Carlo results and
the open symbols show the results obtained by evaluating $\Gamma(p,q)$
perturbatively with the diagrams of
Fig.~\ref{Diagram}. In the perturbative calculations, 
$g(p,q)$ is calculated by using  wave-function renormalizations
$Z(p)$ and $Z(p+q)$ extracted from Monte Carlo data.
As one can see, in the weak-correlation regime,
the perturbative 
calculations are in good agreement with Monte Carlo simulations.
However, when the Hubbard $U$ exceeds $U\approx 4$ ($\sim\frac{W}{2}$),
perturbation theory appears to break down.

QMC calculations of the doping, temperature and $U$ dependence of 
the vertex enhancement and its ${\vv q}$ dependence, which will be 
presented in more detail in a longer paper, give a physical picture for 
both the unexpected increase of the vertex as a function of larger 
$U$-values for small phonon momenta and also the suppression of 
the vertex for large phonon momenta: It is well established, in particular 
in terms of QMC work on the single-particle spectral function of the 
Hubbard model~\cite{groeber}, that the single-particle excitations as
a function of increasing $U$-values undergo (at lower enough 
temperatures, around $\beta=2-3$) a crucial physical transition into a 
strong-correlation regime around $U\approx 6t$: the electronically filled
valence band of width $W$, which is essentially given by the bare
band width $W=8t$, splits into two ``bands". The physical picture 
behind this splitting is the formation of a ``spin-bag" quasiparticle, i.e. of
the bare particle (hole) dressed with a spatially (typically a few lattice
constants) extended spin cloud, which is due to the frustration of 
the local antiferromagnetic order. 
The spin bag moves coherently and ``slowly"
with an energy scale $J={4t^2\over U}$ within the new, strongly 
renormalized quasiparticle band of width $\sim J$. This coherent 
motion couples effectively (with small energy denominators) to
longer wavelength lattice displacements, whose wavelength is typically
longer than the spin-bag ``diameter". On the other hand, there is at larger
$U$-values also an incoherent lower Hubbard band whose higher energy
scale corresponds to the ``rattling around" of the bare particle within
the spin bag~\cite{groeber}. Only large momenta phonons, i.e.
with wavelength smaller than the ``extension" of the spin bag, 
can couple to these incoherent electronic degrees of freedom.
Their coupling is weak because of the combined effects of the 
incoherent motion and the large ($\sim $ scale $W$) energy denominators.

In summary, based on QMC simulations, we have studied
the el-ph vertex function in the two-dimensional Hubbard model.
We find that in the weak-correlation regime,
the effects of the Hubbard interaction U are to {\it suppress} the
ionic el-ph coupling at all phonon momenta,
with backward scattering processes being more strongly suppressed 
than forward ones. On the other hand, in the strong-correlation regime,
the vertex at smaller phonon momentum transfer anomalously
{\it increases} as a function of $U$. We also find that screening is 
the dominant contribution to the vertex corrections at weak correlation,
while at strong correlation the irreducible vertex corrections are crucial.

We would like to acknowledge useful discussions 
with Dr. R. Zeyher and C. Castellani.
The W\"urzburg group would like to acknowledge support by the DFG under Grant
No.~Ha 1537/20-1 and by a
Heisenberg Grant (AR 324/3-1), and by the
Bavaria California Technology Center (BaCaTeC),
the  KONWHIR projects OOPCV and CUHE. DJS acknowledges support from the US 
Department of Energy under Grant \#DOE85-45197. 
The calculations were carried out at the high-performance computing centers
HLRS (Stuttgart) and LRZ (M\"unchen).


\begin{thebibliography}{99}
\bibitem{IFT98}
M. Imada, A. Fujimori, and Y. Tokura, {\sl Rev.~Mod.~Phys.} {\bf 70}, 
1039 (1998).
\bibitem{And02}
P.W.~Anderson, cond-mat/0201429.
\bibitem{raman}
V.G.~Hadjiev, X.J.~Zhou, T.~Strohm, M.~Cardona,
Q.M.~Lin, and C.W.~Chu, {\sl Phys.~Rev.~B} {\bf 58}, 1043 (1998); for a review,
see also M.L.~Kulic, {\sl Physics Reports} {\bf 338}, 1--264 (2000).
\bibitem{iso1}
J.P.~Franck, S.~Harker, and J.H.~Brewer, {\sl Phys.~Rev.~Lett.} {\bf 71}, 
283 (1993).
\bibitem{tunnel1}
D.~Shimada, Y.~Shiina, A.~Mottate, Y.~Ohyagi, and N.~Tsuda,
{\sl Phys.~Rev.~B} {\bf 51}, R16495 (1995).
\bibitem{lanzara}
A.~Lanzara, P.V.~Bogdanov, X.J.~Zhou, S.A.~Keller, D.L.~Feng,
E.D.~Lu, T.~Yoshida, H.~Eisaki, A.~Fujimori, K.~Kishio,
J.-I.~Shimoyama, T.~Noda, S.~Uchida, Z.~Hussain, and Z.-X.~Shen,
{\sl Nature} {\bf 412}, 510 (2001).
\bibitem{kim}
J.H.~Kim, and Z.~Tesanovic, {\sl Phys.~Rev.~Lett.} {\bf 71}, 4218 (1993).
\bibitem{grilli}
M.~Grilli and C.~Castellani, {\sl Phys.~Rev.~B} {\bf 50}, 16880 (1994).
\bibitem{ZK96}
R.~Zeyher and M.L.~Kulic, {\sl Phys.~Rev.~B} {\bf 53}, 2850 (1996).
\bibitem{BS96}
N.~Bulut and D.J.~Scalapino, {\sl Phys.~Rev.~B} {\bf 54}, 14971 (1996).
\bibitem{BSS81}
R.~Blankenbecler, D.J.~Scalapino, and R.L.~Sugar, {\sl Phys.~Rev.~D} 
{\bf 24}, 2278 (1981).
\bibitem{pq} In our convention, unbolded variables denote both
Matsubara frequency and momentum, i.e., $p=(p_0,{\vv p})$ and
$q=(q_0,{\vv q})$. For a static phonon, $q_0$ is set to be zero.
\bibitem{cerruti}
B.~Cerruti, E.~Cappelluti, and L.~Pietronero, cond-mat/0307190.
\bibitem{fint}
The imaginary-time-dependent perturbation \eqref{el-ph}
has a well-defined meaning within a functional-integral formulation,
where it leads to the linear-response result \eqref{four}.
Details will be given elsewhere.
\bibitem{ref1} This method could, of course, be extended to take into
account phonon vertex corrections, but here we are interested in
understanding the effects of the Hubbard interaction $U$ on the vertex.
\bibitem{ref2}Note that $E_{\vv p}$ is a complex variable due to 
lifetime effects. For simplicity, we have assumed ${\rm Im}E_{\vv p}<< 
p_{0}$ and extracted the wave-function renormalization $Z(p)$ from
the equation ${\rm Im} \frac{1}{G(p)} = Z(p)\,p_0$.
\bibitem{holdmft}
A. Deppeler and A.J.~Millis, {\sl Phys.~Rev.~B} {\bf 65}, 100301 (2002);
{\it ibid.} {\bf 65}, 224301 (2002); S. Blawid, A. Deppeler, and A.J.~Millis,
{\it ibid.} {\bf 67}, 165105 (2003); J.K.~Freericks, {\it ibid.} {\bf 50},
403 (1994); J.K.~Freericks and M.~Jarrell, {\it ibid.} {\bf 50}, 6939 (1994).
\bibitem{groeber}
C.~Gr\"ober, R.~Eder, and W.~Hanke, {\sl Phys.~Rev.~B} {\bf 62}, 
4336 (2000), and references therein.
\end{thebibliography}
\end{document}